\documentclass[journal]{IEEEtran}
\ifCLASSINFOpdf
\else
\fi

\usepackage{amsmath} 
\usepackage{amssymb} 
\usepackage{bm}      
\usepackage{algorithmic}    
\usepackage[linesnumbered,ruled]{algorithm2e} 
\usepackage{cite}
\usepackage{graphicx}  
\usepackage{subfigure}  
\usepackage{multirow}
\usepackage{url}
\usepackage{color}
\usepackage{booktabs} 
\usepackage{hyperref} 

\begin{document}
%
\title{Learning Deep Direct-Path Relative Transfer Function for Binaural Sound Source Localization}
%
%
%

\author{Bing~Yang, Hong~Liu, and Xiaofei~Li
\thanks{This work is supported by National Natural Science Foundation of China (No.\,62073004), Science and Technology Plan of Shenzhen (No.\,JCYJ20200109140410340). }
\thanks{B. Yang is with the Key Laboratory of Machine Perception, Peking University, Shenzhen Graduate School, Beijing 100871, China (e-mail: bingyang@sz.pku.edu.cn).}
\thanks{Corresponding author: H. Liu is with the Key Laboratory of Machine Perception, Peking University, Shenzhen Graduate School, Beijing 100871, China (e-mail: hongliu@pku.edu.cn).}
\thanks{X. Li is with Westlake University \& Westlake Institute for Advanced Study, Hangzhou 310024, Zhejiang, China (e-mail: lixiaofei@westlake.edu.cn).}
}

\maketitle

\begin{abstract}
Direct-path relative transfer function (DP-RTF) refers to the ratio between the direct-path acoustic transfer functions of two microphone channels. Though DP-RTF fully encodes the sound spatial cues and serves as a reliable localization feature, it is often erroneously estimated in the presence of noise and reverberation. This paper proposes to learn DP-RTF with deep neural networks for robust binaural sound source localization. A DP-RTF learning network is designed to regress the binaural sensor signals to a real-valued representation of DP-RTF. It consists of a branched convolutional neural network module to separately extract the inter-channel magnitude and phase patterns, and a convolutional recurrent neural network module for joint feature learning. To better explore the speech spectra to aid the DP-RTF estimation, a monaural speech enhancement network is used to recover the direct-path spectrograms from the noisy ones. The enhanced spectrograms are stacked onto the noisy spectrograms to act as the input of the DP-RTF learning network. We train one unique DP-RTF learning network using many different binaural arrays to enable the generalization of DP-RTF learning across arrays. This way avoids time-consuming training data collection and network retraining for a new array, which is very useful in practical application. Experimental results on both simulated and real-world data show the effectiveness of the proposed method for direction of arrival (DOA) estimation in the noisy and reverberant environment, and a good generalization ability to unseen binaural arrays.
\end{abstract}


%
\IEEEpeerreviewmaketitle

\section{Introduction}
%
%
%
%
\IEEEPARstart{S}{ound} source localization using microphone arrays has been investigated intensively by many researchers in the last decades, due to its importance and wide application in teleconferencing, robot audition and hearing aids. It provides important characteristics of sound sources that can boost a variety of signal processing tasks such as speech enhancement, sound source separation and speaker recognition \cite{SESSOverview17, SESSOverview18}. Many researchers adopt a dual-stage approach which consists of localization feature extraction and feature-to-location mapping \cite{SG11,YBICASSP17,YBTASLP19}. Under the dual-stage framework, reliable  feature extraction is extremely important for robust sound source localization.

The commonly used spatial cues to represent the source location are the time and intensity difference between dual-microphone signals. The time difference features are more reliable on low frequency due to a reduced possibility of spatial aliasing, which include inter-channel time difference (ITD) \cite{ITDIID10}, inter-channel phase difference (IPD) \cite{IPD10}, generalized cross-correlation (GCC) function \cite{GCC76}, etc. The intensity difference feature, often referred to inter-channel intensity difference (IID) \cite{ITDIID10}, is more reliable on high frequency where the head shadowing effect exists. Since IID is less discriminative for free-field microphones and can be easily contaminated by acoustic interferences, it is hardly used for localization solely. The complementarity of the two types of difference features contributes the fusion of time and intensity difference information. A typical fused feature is relative transfer function (RTF) \cite{RTF15, RTF18} which encodes time and intensity difference in its argument and magnitude respectively.

The aforementioned localization features can be easily estimated under a noise-free and anechoic condition. In practical acoustic scenes, microphone signals are composed of the direct-path propagated source signal, the sound waves reflected by environment and the ambient noise. Since reflections cause an overlap-masking effect or a coloration of the anechoic signal \cite{Derev10} and noises add uncertain acoustic distortion to microphone signals, the accuracy of feature estimation will be degraded in the presence of noise and reverberation, which further leads to a significant drop on the localization performance.
Considering the fact that only the time and intensity information corresponding to the direct-path sound propagation are reliable for source localization, this work dedicates to extract direct-path RTF (DP-RTF) \cite{LXF16,LXF19,YBCAAI21} feature for robust sound source localization. The deep neural network (DNN) is taken as the tool due to its strong modelling ability. A DP-RTF learning framework that embeds the sensor signals to a low-dimensional localization feature space is designed, which disentangles the localization cues from other factors including source signals, noise, reverberation, etc. The DP-RTF learning based localization method takes full use of the spatial and spectral cues, which is demonstrated to perform better than several other methods on both simulated and real-world data in the noisy and reverberant environment. The proposed method is an extended version of our previous work \cite{YBICASSP21}, which has the following contributions.

\subsubsection{A novel DP-RTF learning network}
The complex DP-RTF cannot be directly learned by the real-valued DNN. In order to fit the real-valued network, the complex DP-RTF is changed into a real-valued representation, namely a concatenation of the IID and the sinusoidal functions of IPD, which keeps equivalent information as the complex DP-RTF. A convolutional recurrent neural network (CRNN) is designed for DP-RTF learning. The log-magnitude and phase spectrograms of binaural signals are taken as network input. Especially, considering the heterogeneity of magnitude and phase information, the log-magnitude and phase spectrograms are processed by two separate convolutional neural network (CNN) branches to respectively extract the spatial representations of magnitude and phase.  In addition, to give more attention to the reliable spatial features, one extra CNN branch taking as input the log-magnitude spectrogram is designed to estimate a gate/weight/mask spectrogram, which is then applied to the spatial representations obtained by the other two CNN branches. The masked spatial representations are further passed to the following CRNN module to predict the DP-RTF representation. The network is trained with the mean squared error (MSE) loss between the predicted DP-RTF and the ground truths. The DP-RTF prediction can be used to estimate the direction of arrival (DOA) of the source by matching the predicted DP-RTF with the ground truths of candidate directions.

\subsubsection{Leveraging monaural speech enhancement  to improve the robustness of DP-RTF estimation}
The DP-RTF learning network proposed above focuses on extracting the inter-channel DP-RTF information, while the information of monaural speech spectra is not especially explored. Monaural speech enhancement techniques \cite{SESSOverview18} learn the speech spectral pattern to recover the clean speech from the contaminated one.  The enhanced speech would be definitely helpful for DP-RTF estimation. In this work, we adopt the network architecture of the monaural speech enhancement method in \cite{FullSub21}. This enhancement method is modified to recover the clean direct-path magnitude and phase spectrograms from the contaminated ones, instead of recovering the noise-free signals. Then, the enhanced  spectrograms  together with the contaminated  spectrograms are utilized for DP-RTF learning.

\subsubsection{Generalization to unseen binaural configurations}
The DP-RTF of different binaural configurations (w.r.t the shape of torso and head) are normally different. In real-world applications, to train a localization network for a new binaural configuration, one need to either collect a large amount of data or measure the head-related impulse responses (HRIRs) of all 3D directions for data simulation, which are both time-consuming. In this work, we propose to train one unique network with many different binaural configurations, so that hopefully the network can be generalized to unseen binaural configurations. Experimental results show that the proposed method achieves superior generalization performance, as the training configurations can cover a large range of DP-RTF distributions, and the network is able to learn the interpolation between binaural configurations.

The rest of this paper is organized as follows. Section II overviews the related works in the literature. Section III formulates the DP-RTF prediction problem for sound source localization. Section IV details the proposed DP-RTF learning network. Experiments and discussions with simulated and real-world data are presented in Section V, and conclusions are drawn in Section VI.

\section{Related Works}

\begin{table*}[]
    \centering
    \caption{Deep learning based sound source localization}
    \label{tab:methodoverview}
      \renewcommand\arraystretch{1.05}
      \tabcolsep0.02in
    \begin{tabular}{cccccccccccccccccccccc}
        \hline
        \hline
        \multirow{2}{*}{Category} &\multirow{2}{*}{Approach} &\multirow{2}{*}{Input} &\multirow{2}{*}{Output} &\multirow{2}{*}{Network} &Enhancement against &Generalization \\
        & & & & &noise/reverberation &to unseen array  \\
        \hline
        Feature-to-location
        &\cite{DNN15}   &GCC &Location  identity  &FC &No &No\\
        &\cite{DNN16}   &Eigenvectors of spatial correlation matrix &Location identity  &FC &No &No\\
        &\cite{CCF17}   &GCC, IID       &Location  identity  &FC  &No &Yes\\
        &\cite{CNN19}   &Phase spectrum &Location identity  &CNN &No &No\\
        &\cite{Mask_DOA19} &Magnitude (single channel) and phase spectrum &Location identity  &CNN &Yes &No\\
        &\cite{CRNN_intensity19} &Intensity vector  (ambisonics format) &Location identity  &CRNN &No &No \\
        &\cite{SCL_CNN20} &Spatial pseudo-spectrum &Location identity  &CNN &No &No
        &\vspace{0.12cm}\\
        Signal-to-location
        &\cite{ETE_Ma19}&Sensor signals &Location identity  &CNN &No &No\\
        &\cite{CRNN18}  &Magnitude and phase spectrum (ambisonics format) &Location identity
        &CRNN &No &No\\
        &\cite{CRNN19}  &Magnitude and phase spectrum (ambisonics format)  &Location coordinate  &CRNN &No &No
        &\vspace{0.12cm}\\
        Feature-to-feature
        &\cite{Mask_TDOA19} &Magnitude (single channel) spectrum, GCC&ITD &RNN &Yes &No \\
        &\cite{Embed19} &Sine and cosine of IPD, IID  &Embedding &FC &Yes &No\\
        &\cite{IPD19}   &Sine and cosine of IPD &Sine and cosine of IPD  &FC  &Yes &No\\
        &\cite{MaskIPD21}   & Magnitude spectrum, sine and cosine of IPD  &Sine and cosine of IPD  &RNN, FC  &Yes &No\\
        &\cite{YBCAAI21}&Sine and cosine of IPD, IID  &DP-RTF &FC &Yes &No\\
        \hline
        Signal-to-feature
        &Proposed &Magnitude and phase spectrum &DP-RTF &CRNN &Yes &Yes\\
        \hline
        \hline
    \end{tabular}
\end{table*}

\subsection{Deep Learning Based Sound Source Localization}

With the development of deep learning techniques, lots of localization works
\cite{DNN15, DNN16, CCF17, CNN19, Mask_DOA19, CRNN_intensity19,SCL_CNN20,ETE_Ma19, CRNN18, CRNN19,Mask_TDOA19, Embed19, IPD19, MaskIPD21, YBCAAI21} are built in a supervised manner. A summary of the recent deep learning based sound source localization methods is presented in Table~\ref{tab:methodoverview}. They treat the localization task as either a classification or a regression problem. Compared with the conventional unsupervised methods, these works are data-driven and hence can better adapt to various acoustic conditions that present in the training data.

According to the role of the deep learning model plays, the previous works are classified into three categories, namely feature-to-location \cite{DNN15, DNN16, CCF17, CNN19, Mask_DOA19, CRNN_intensity19,SCL_CNN20}, signal-to-location \cite{ETE_Ma19, CRNN18, CRNN19}, and feature-to-feature \cite{Mask_TDOA19, Embed19, IPD19, MaskIPD21, YBCAAI21} methods.
Feature-to-location and signal-to-location methods are able to learn the non-linear functions that map features or signals to source location. Feature-to-feature methods provide a simple and effective way to recover the core localization features from the distorted features.
As shown in Table~\ref{tab:methodoverview}, the utilized network architectures for sound source localization include fully connected (FC) neural network \cite{DNN15, DNN16, CCF17, Embed19, IPD19, MaskIPD21, YBCAAI21}, CNN \cite{CNN19, Mask_DOA19, SCL_CNN20}, recurrent neural network (RNN) \cite{Mask_TDOA19} and CRNN \cite{CRNN_intensity19,CRNN18, CRNN19}.
As for the network input, the spatial features used for training include inter-channel difference features such as IPD \cite{Embed19, IPD19, MaskIPD21, YBCAAI21}, IID \cite{CCF17, Embed19, YBCAAI21}, eigenvectors of spatial correlation matrix \cite{DNN16}, intensity vector\cite{CRNN_intensity19}, GCC \cite{DNN15, CCF17, Mask_TDOA19}, and the spatial spectra such as spatial pseudo-spectrum \cite{SCL_CNN20}. The  magnitude spectrum \cite{Mask_DOA19,  Mask_TDOA19, MaskIPD21} can be also fed together with the spatial features to the network, but cannot be used solely. The signal taken as input can be the time-domain signal \cite{ETE_Ma19}, or the magnitude and phase of short-time Fourier transform (STFT) coefficients \cite{CRNN18, CRNN19}, which contains full spectral and spatial information.
Taking as input the signal is expected to better learn the spatial cues, as long as the network is well designed. The source location class \cite{DNN15, DNN16, CCF17,CNN19, Mask_DOA19, SCL_CNN20, ETE_Ma19, CRNN18, CRNN_intensity19} is taken as output in the classification framework, while the location coordinate \cite{CRNN19} or the localization feature \cite{Mask_TDOA19, Embed19, IPD19, MaskIPD21, YBCAAI21} in the regression framework.

Most data-driven approaches train the localization network for single array, and apply the network to the same array. The trained model will not perform well on other unseen arrays, because the mappings from localization features to source locations differ from one array to another. Few works concern the generalization to unseen arrays. Ma \emph{et al}. \cite{CCF17} used the multi-condition training (MCT) to deal with the head mismatch between training and test. Wang \emph{et al}. \cite{Head20} studied the binaural localization in the mismatched head-related transfer function (HRTF) condition and proposed a data-efficient method based on DNN and clustering. These works learn a common network for different array configurations to implement the feature-to-location mapping. The mapping function is determined by the direct-path acoustic propagation model, such as the far-field plane wave model for regular free-filed microphone array, the HRIRs for binaural audition, etc. As the feature-to-location mapping functions of two arrays may be similar for different source locations, the tolerance to multiple arrays will increase the intra-class variation and make the trained network confused with the two locations. Though the works \cite{SCL_CNN20,IPD19, MaskIPD21, YBCAAI21} did not conduct the array generalization experiments, they can convert the array-related features to array-independent features/locations if the direct-path acoustic propagation model is given beforehand. Further investigation is required about how to use DNN to deal with the array generalization problem.

Different from the existing three types of approaches, the proposed method uses a special signal-to-feature framework. Taking as network input the microphone signals allows to make full use of the redundant spectral and spatial information to disentangle the intrinsic localization cues of the sound source from other factors including noise and reverberation.
Accordingly, we have specially designed branched layers to extract the two types of information. DP-RTF is taken as network output, which can be trivially mapped to the source location as long as the microphone topology (or HRIRs for binaural microphones) is known. Importantly, DP-RTF can be predicted from the microphone signals regardless of the binaural array configurations, and thence the DP-RTF learning network can be directly applied to unseen arrays.

\subsection{Enhancement of Signal Spectra and Localization Features}

Acoustic interferences are inevitable in real-world acoustic scenes. Ambient noise and room reverberation often add uncertain acoustic distortion to sensor signals or localization features, which degrades the sound source localization performance. Two types of methods are designed to improve the robustness of localization against noise and reverberation, namely weighting \cite{Mask_CNN17,Mask_RNN19,MaskIPD21,Mask_Unet20,Mask_DOA19,Mask_TDOA19,Mask_ATT20,CT08,DPD14,Griffin13-SCL} and enhancement \cite{LXF16,Embed19,IPD19,YBCAAI21,Mask_TDOA19,MaskIPD21} methods.

The weighting methods reduce the effect of acoustic interferences by highlighting the time-frequency (TF) regions dominated by direct sound, and using the TF-weighted signals or features for further localization. The TF weight/mask can be predicted by unsupervised methods such as coherence test \cite{CT08}, direct path dominance (DPD) test \cite{DPD14}, single source confidence measure \cite{Griffin13-SCL}, or the DNN models \cite{Mask_CNN17, Mask_RNN19,MaskIPD21, Mask_Unet20, Mask_DOA19, Mask_TDOA19, Mask_ATT20}. The mask with binary values may cause the selection error, i.e., miss-detections and false-detections, of TF bins, which will lead to localization error. In contrast to the weighting methods, the enhancement methods aim to directly remove the acoustic interferences and retain direct-path information.
Li \emph{et al}. \cite{LXF16,LXF19} used a convolutive transfer function model and an inter-frame spectral subtraction algorithm to separately suppress reverberation and noise in order to identify the DP-RTF. Tang \emph{et al}. \cite{Embed19} designed a siamese architecture to learn a low-dimensional representation of the localization cues that is consistent with the source location. Pak \emph{et al}. \cite{IPD19} and Cheng \emph{et al}. \cite{MaskIPD21} trained DNN models to enhance the interference-contaminated IPD on the sinusoidal tracks.

Despite the progress of these research, most above works exploit partial of the spectral and spatial information to improve the robustness of localization. In this work, a DP-RTF learning method that makes full use of both spatial and spectral patterns is proposed for robust DOA estimation. In addition to a CRNN module for joint magnitude and phase feature learning, extra CNN branches are used to extract and highlight feature patterns. Besides, the monaural speech enhancement technique is employed to recover the direct-path spectral patterns of the sound source before DP-RTF learning. Different from most previous works that only use DNN to learn spectral or spatial cues, this work uses DNN to learn both.

\section{Problem Formulation}
\begin{figure}[t]
  \centering
  \includegraphics[width=0.92\linewidth]{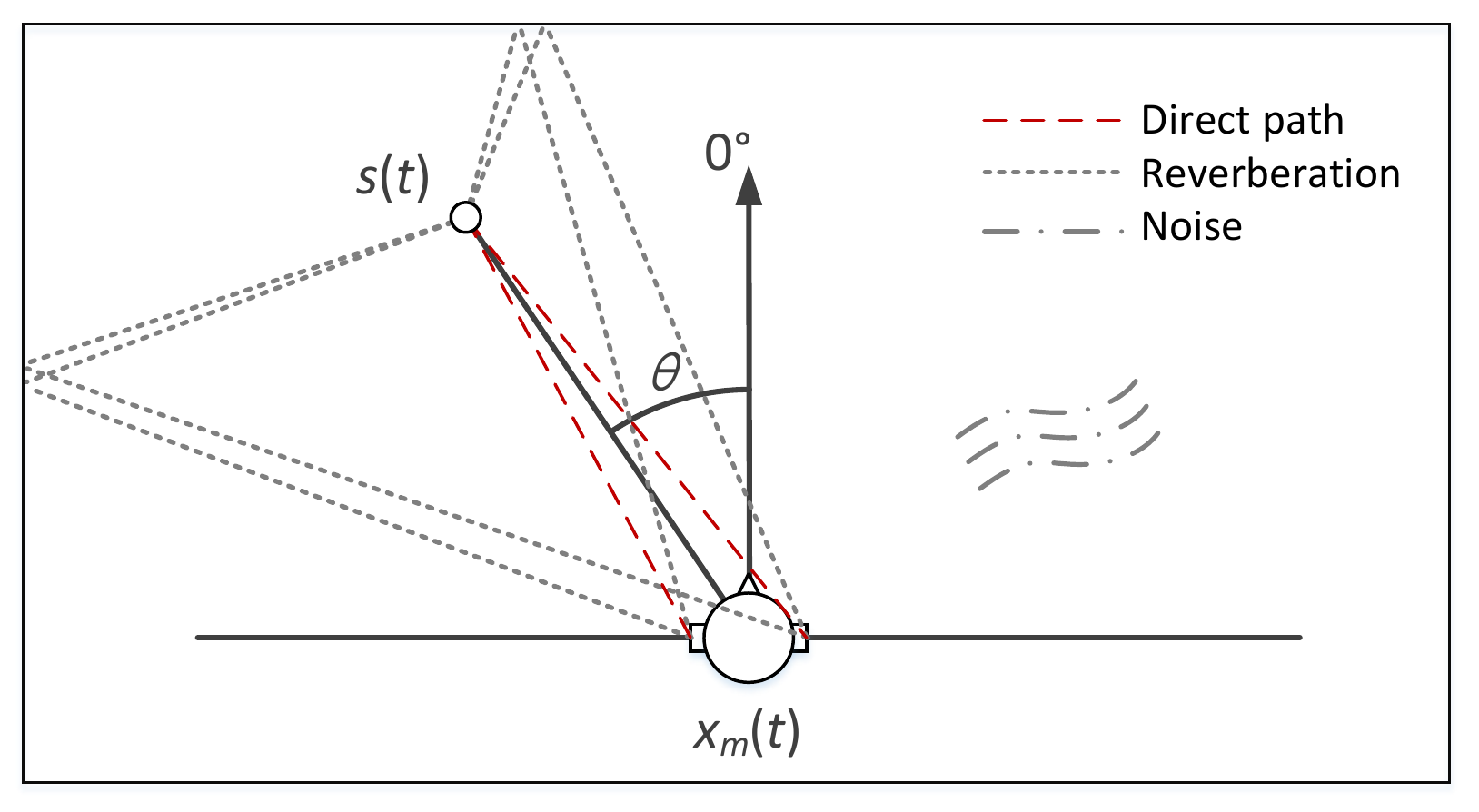}
  \caption{Illustration of the signal model for the considered acoustic scene. }
  \label{fig:meth_scene}
\end{figure}
We consider a single sound source observed by binaural microphone pair, equipped in the dual ears of a dummy head, in an enclosed environment with additive ambient noise as shown in Fig.~\ref{fig:meth_scene}. The signal received by the $m$-th microphone is denoted as
\begin{equation}
    x_m(t) = h_{m}(t, \theta) * s(t) + v_m(t),
  \label{eq_model_T}
\end{equation}
where $t \in  [1, T]$ represents the time sample index, $m \in [1,2]$ is the microphone index, $\theta$ denotes the horizontal DOA of the source, $s(t)$ denotes the source signal, $v_m(t)$ denotes the received noise signal at the $m$-th microphone, and $h_{m}(t, \theta)$ is the acoustic impulse response from the source at  $\theta$ to the $m$-th microphone. Here, $*$ denotes the convolution operation. Applying the STFT to Eq. \eqref{eq_model_T}, the microphone signal is expressed in the TF domain as
\begin{equation}
    X_m(n,f) =  H_m(f, \theta) S(n,f)  + V_m(n,f),
    \label{eq_model_TF}
\end{equation}
where $n \in  [1, N]$ represents the time frame index, $f \in  [1, F]$ represents the frequency index.  $N$ and $F$ are the number of frames and frequencies, respectively. Here, $X_m(n,f)$, $S(n,f)$ and $V_m(n,f)$ represent the microphone, source and noise signals in the TF domain, respectively.
The acoustic transfer function $H_m(f,\theta)$ is the Fourier transform  of $h_{m}(t, \theta)$. It is assumed to be time-invariant during the time period of interest, and hence it is not a function of time frame index. The acoustic transfer function involves the direct and reflected propagation paths of the sound source to the microphones, i.e.,
\begin{equation}
     H_m(f,\theta) = H_m^{\rm{d}}(f,\theta)+H_m^{\rm{r}}(f,\theta),
    \label{eq_H}
\end{equation}
where $H_m^{\rm{d}}(f,\theta)$ and $H_m^{\rm{r}}(f,\theta)$ denote the acoustic transfer functions of direct path and reverberation (note that, it actually includes both early reflections and late reverberation), respectively. The direct-path relative transfer function (DP-RTF) \cite{LXF16,LXF19} is defined as the ratio between the two direct-path acoustic transfer functions, namely
\begin{equation}
     R^{\rm{d}}(f,\theta) =
     \frac{ H_2^{\rm{d}}(f,\theta)}{ H_1^{\rm{d}}(f,\theta)}.
     \label{eq_RTF_DP}
\end{equation}

The difference between the direct-path signals of two microphone channels fully encodes the source location, which should be independent of other acoustic factors such as room characteristics, noise signals and source signals.
Hence, this work aims to use DNN to embed the microphone signals $X_m(n,f)$ recorded in the presence of noise and reverberation into a target space that only preserves the acoustic characteristics related to DP-RTF, such that sound source localization can be performed by directly matching the predicted DP-RTF representation with the ground truths of candidate directions.

\section{DP-RTF Learning for Sound source localization}
In this section, we first give an overview of the DP-RTF learning based DOA estimation framework. Then, the real-valued DP-RTF representation is defined. Finally, the network architecture is described in details.

\subsection{Overview of DOA Estimation Framework}

\begin{figure}[t]
  \centering
  \includegraphics[width=1\linewidth]{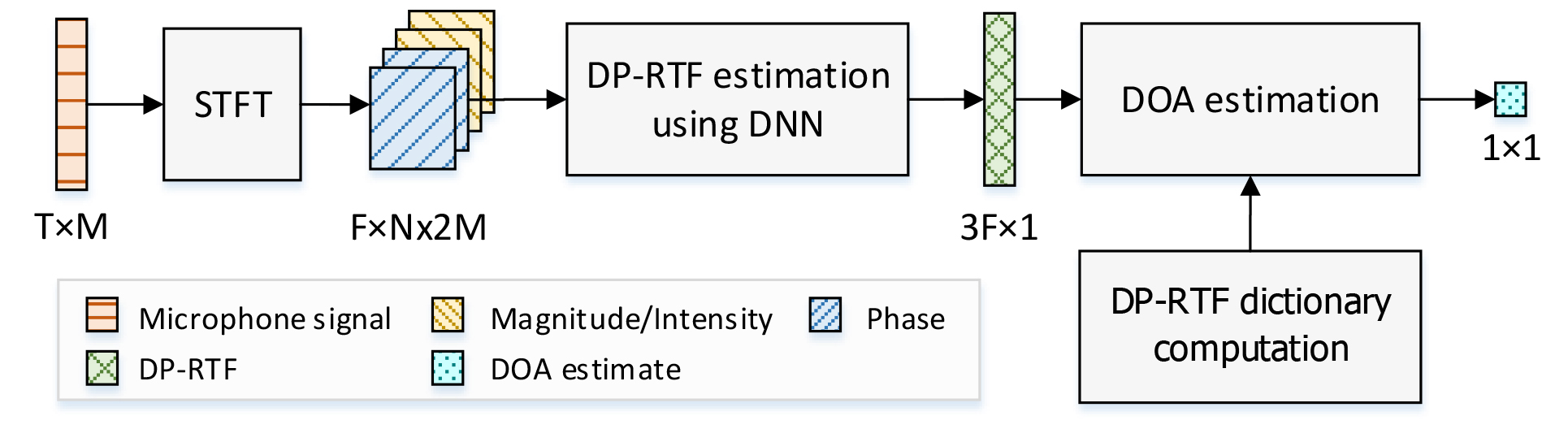}
  \caption{Block diagram of binaural sound source localization based on DP-RTF learning. }
  \label{fig:meth_flowchart}
\end{figure}

\begin{figure*}[t]
  \centering
  \includegraphics[width=0.86\linewidth]{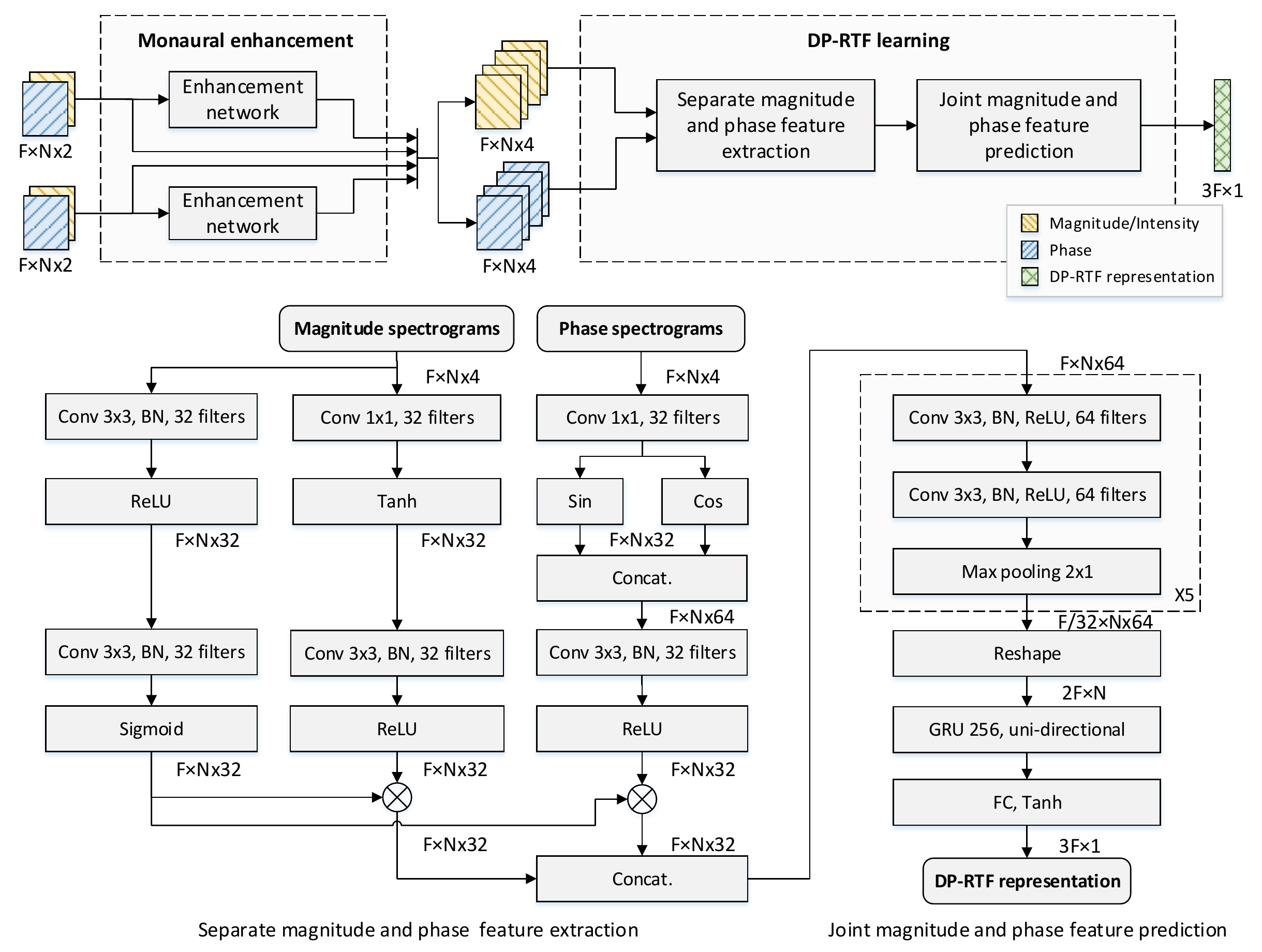}
  \caption{Model architecture for DP-RTF estimation. The model consists of two parts, namely the monaural enhancement network and the DP-RTF learning network.}
  \label{fig:meth_dnn}
\end{figure*}

The block diagram of the DP-RTF learning based binaural sound source localization method is shown in Fig.~\ref{fig:meth_flowchart}. The dictionary of DP-RTFs is constructed using the HRIRs of all candidate directions. We transform the microphone signals into the TF domain, and then estimate DP-RTF using the DNN model shown in Fig.~\ref{fig:meth_dnn}. The DNN model consists of a monaural enhancement network and a DP-RTF learning network (see details in Section IV-C and IV-D). The monaural speech enhancement method is used to enhance the direct-path spectrograms. With the dual-channel contaminated and enhanced spectrograms, DP-RTF is predicted by a separate-to-joint feature learning process. Finally, the DOA of the sound source is estimated by looking up the DP-RTF dictionary (taking the direction that minimizes the dictionary matching result), i.e.,
\begin{equation}
    \hat{\theta} = \arg\min_{\theta}||\hat{\mathbf{r}}  - \mathbf{r}(\theta)||^2 ,
    \label{eq_SSL}
    \vspace{-0.1cm}
\end{equation}
where $\hat{\mathbf{r}}$ is the DP-RTF prediction, and $\mathbf{r}(\theta)$ is the lookup in DP-RTF dictionary (see details in Section IV-B).

\subsection{Real-valued DP-RTF Representation}

In theory, the direct-path acoustic transfer function, more specifically the HRTF in the binaural localization context, can be expressed as
\begin{equation}
     H_m^{\rm{d}}(f,\theta) = \alpha_{m}(f,\theta)e^{-j\omega_f\tau_{m}(\theta)},
    \label{eq_H_DP}
\end{equation}
where $\omega_f$ denotes the angular frequency of the $f$-th frequency, and $\alpha_{m}(f,\theta)$ and $\tau_{m}(\theta)$ are the propagation attenuation factor and the time of arrival from the source to the $m$-th microphone, respectively. Substituting it into Eq.~\eqref{eq_RTF_DP}, DP-RTF can be rewritten as
\begin{equation}
     R^{\rm{d}}(f,\theta) = \frac{\alpha_{2}(f,\theta)}{\alpha_{1}(f,\theta)}e^{-j\omega_f\left(\tau_{2}(\theta)-\tau_{1}(\theta)\right)}.
    \label{eq_RTF_DP2}
\end{equation}
DP-RTF encodes IID and IPD in its magnitude and argument respectively. Considering the complex-valued DP-RTF cannot be directly processed by the real-valued DNN, the DP-RTF representation is carefully designed without information loss.

The phase-magnitude decomposition is used to map the complex values to real ones. The phase of DP-RTF is exactly the IPD, which is denoted as
\begin{equation}
      \Delta P(f,\theta) = \angle{R}^{\rm{d}}(f,\theta),
      \label{eq_IPD}
\end{equation}
where $\angle$ is the phase operator of complex numbers. The IPD is in the range from $-\pi$ to $\pi$. It tends to be periodically wrapped with the increasing of frequency or time difference, and discrete when $\omega_f(\tau_{2}(\theta)-\tau_{1}(\theta))$ reaches $\pi+2i\pi$ with an integer $i$. To avoid the phase wrapping ambiguity and retain the local continuity, the sinusoidal functions of IPD are used instead, namely $\sin\Delta P(f,\theta)$ and $\cos\Delta P(f,\theta)$, which are continuous in [-1,1].  Since DP-RTF is defined as a ratio between two values,  the magnitude of DP-RTF is asymmetrical with respect to the broadside direction of the two microphones. Instead, we transform the magnitude into log domain, as the IID defines
\begin{equation}
     \Delta I(f,\theta) = \frac{20\log_{10}\left|{R}^{\rm{d}}(f,\theta)\right|}{\Delta I_{\rm{max}}},
      \label{eq_IID}
\end{equation}
where $|\cdot|$ denotes the magnitude of complex numbers. $\Delta I_{\rm{max}}$ is an empirically-set maximum value of IID, which is used for normalizing the IID into the range of  [-1,1] to balance the contribution of IID and IPD.
Accordingly, the real-valued DP-RTF representation is defined as a concatenation of normalized IID, and the sinusoidal functions of IPD over all frequencies, i.e.,
\begin{equation}
\begin{aligned}
    {\mathbf{r}(\theta)} =&
    [\Delta I(1,\theta),\ldots,\Delta I(F,\theta), \\
    &\sin\Delta P(1,\theta),\ldots,\sin\Delta P(F,\theta), \\
    &\cos\Delta P(1,\theta), \ldots, \cos\Delta P(F,\theta)]^T \in  \mathbb{R}^{3F\times1},
    \label{eq_DP_RTF_vec}
\end{aligned}
\end{equation}
where $(\cdot)^T$ denotes vector transpose.

\subsection{Monaural Enhancement}

Deep learning has been widely used for monaural speech enhancement \cite{SESSOverview18}. Deep monaural speech enhancement recovers the clean magnitude spectrogram by learning the structured magnitude spectral pattern of speech. The magnitude/intensity spectrum serves as an important cue to indicate the direct-path dominance of TF regions. In addition, the inter-channel magnitude/intensity difference plays an especially important role for binaural localization, as the intensity difference of binaural signals can reflect the torso/head shadow effect of signal propagation. In order to promote the localization performance, the recently proposed FullSubNet \cite{FullSub21} is adopted to predict the complex ideal ratio mask and enhance the complex speech spectrograms. Accounting for the following DP-RTF learning, the clean direct-path sound is taken as the target signal, which means both noise reduction and dereverberation are conducted. During test, the binaural microphone signals are separately enhanced using the same monaural speech enhancement network. The enhanced signals are used in the following DP-RTF learning step. Note that, although the enhanced phase spectrograms can only slightly improve the DP-RTF learning performance, it is still used.

\subsection{DP-RTF Learning}

Sound source localization can be treated as DOA classification or feature regression problems, since the direction label and localization cues can be transformed to each other if the direct-path acoustic propagation model is given. However, the mapping from localization cues to DOA for different array configurations can be hardly modeled by one common network, when treating localization as a classification problem. Hence, the DNN is used to regress the direct-path localization cues/features from sensor signals, and the feature-to-location mapping is implemented according to the direct-path acoustic propagation model.

The magnitude and phase spectrograms of binaural signals are taken as the network input, from which the inter-channel localization features can be extracted. Meanwhile the spectral characteristic of the sound signal is also presented in such network input, which can aid the extraction of localization features. Both original and enhanced binaural signals are used, as they are complementary in the sense that the original signals are noisy but less speech-distorted, while the enhanced signals are less noisy but possibly speech-distorted. The input spectrograms are fed into three separate network branches to respectively learn the magnitude features, phase features and mask/weight spectrograms, which are then passed to a joint learning process. We use the MSE loss to train the DP-RTF learning network, and the training target is the ground truth DP-RTF that can be precomputed using the HRIRs following Eq.~\eqref{eq_DP_RTF_vec}.

\subsubsection{CNN branches for separate feature learning}
The separate learning process contains three branches, namely magnitude feature extraction, phase feature extraction and mask estimation. The magnitude spectrogram is transformed to the logarithm domain to keep consistent with the log-IID defined in Eq.~\eqref{eq_IID}. Each of log-magnitude and phase spectrograms of both contaminated and enhanced binaural signals is taken as one feature map of convolutional layers. Due to the heterogeneity of magnitude and phase information, they are processed by two separate branches. The log-magnitude spectrograms are fed into a convolutional layer with 32 1 $\times$ 1 kernels and a tanh activation function to extract the inter-channel intensity features for each frequency and time frame. It is then followed by a convolutional layer with 64 3 $\times$ 3 kernels, a batch normalization (BN) and a rectified linear unit (ReLU) activation function, in order to guarantee the learned magnitude features to be non-negative.
Similarly, taking the phase spectrograms as input, a convolutional layer with 32 1 $\times$ 1 kernels is applied to extract the inter-channel phase features for each frequency and time frame.
The phase features are closely related to IPD, which are thus further activated by sine and cosine functions as is done in Eq.~\eqref{eq_DP_RTF_vec}. The phase branch is also followed by a convolutional layer with 64 3 $\times$ 3 kernels, a BN and a ReLU activation function. The mask/weight branch processes the log-magnitude spectrograms with a convolutional layer with 64 3 $\times$ 3 kernels, a BN and a ReLU activation function, and then a convolutional layer with 64 3 $\times$ 3 kernels, a BN and a Sigmoid activation function. The output of mask branch, which reflects the significance of TF bins for DP-RTF learning, is element-wise multiplied with the learned magnitude and phase features.

\subsubsection{CRNN module for joint feature learning}
The intensity-based and phase-based features produced by the separate process are concatenated along the feature map dimension. To capture the relationship between the intensity and phase information,
the concatenated features are passed to 10 convolutional modules with each consisting of a convolutional layer followed by a BN and a ReLU activation function. These convolutional layers are with 64  3 $\times$ 3 kernels. After each two convolutional modules a max pooling is used to compress the frequency dimension. With multiple-frame features outputted by CNN, one-layer uni-directional gated recurrent unit (GRU) with 256 hidden units is utilized to capture the long-distance temporal context information and output a single-frame feature. This is followed by a FC layer to predict DP-RTF, and then an activation of tanh function to fit each DP-RTF element into the range from -1 to 1.

\section{Experiments and Discussions}

\begin{table*}[t]
  \caption{Room configuration for training and test data}
  \label{tab:room}
  \centering
  \renewcommand\arraystretch{1.0}
  \tabcolsep0.11in
  \begin{tabular}{ccccccccccc}
    \hline
    \hline
    Dataset &Room size [m$^3$]  &Array center [m] &Source-to-array distance [m] &RT$_{60}$ [s] &SNR [dB] & Head subject ID\\
    \hline
    \multirow{10}{*}{Training}
    &6.7$\times$9.0$\times$4.6  &(2.5, 4.0, 1.8) &3.0, 3.6 &0: 0.28: 0.84  &-5: 5: 20 &58, 59, 60\\
    &7.0$\times$8.0$\times$5.0  &(3.0, 3.5, 1.7) &1.5, 2, 2.5, 3, 3.4 &0: 0.17: 0.85  &-5: 5: 20 &48, 50, 51\\
    &8.0$\times$6.5$\times$3.6  &(4.0, 3.3, 1.75) &1.5, 2.9 &0: 0.27: 0.81 &-5: 5: 20 &10, 28, 124\\
    &7.0$\times$7.0$\times$4.0  &(3.5, 3.0, 1.6) &1.0, 2.0, 3.0 &0: 0.18: 0.90  &-5: 5: 20 &11, 12, 165\\
    &5.3$\times$8.0$\times$3.8  &(2.4, 1.4, 1.3) &1.8, 2.4 &0: 0.23: 0.92  &-5: 5: 20 &147, 148, 152\\
    &5.0$\times$6.0$\times$2.8  &(2.0, 3.1, 1.45) &0.5, 1.5, 2.5 &0: 0.21: 0.84  &-5: 5: 20 &44, 127, 156\\
    &4.5$\times$6.0$\times$3.1  &(2.0, 3.0, 1.67) &0.8, 2.2 &0: 0.26: 0.78  &-5: 5: 20 &15, 17, 18\\
    &4.0$\times$5.5$\times$3.0  &(2.5, 2.5, 1.4) &0.5, 1.0 &0: 0.25: 0.75  &-5: 5: 20 &134, 135, 137\\
    &5.0$\times$3.2$\times$2.9  &(2.0, 1.5, 1.2) &0.6, 1.2 &0: 0.19: 0.95  &-5: 5: 20 &158, 162, 163\\
    &3.8$\times$3.0$\times$2.5  &(1.2, 1.45, 1.55) &0.75, 1.25 &0: 0.30: 0.90  &-5: 5: 20 &153, 154, 155\\
    \hline
    \multirow{2}{*}{Validation}
    &6.0$\times$6.0$\times$3.5  &(3.5, 3.0, 1.65) &1.75, 2.25 &0: 0.22: 0.88  &-5: 5: 20 &61, 65, 119\\
    &4.0$\times$6.0$\times$3.2  &(2.0, 3.5, 1.35) &0.75, 1.25 &0: 0.24: 0.72  &-5: 5: 20 &126, 131, 133\\
    \hline
    \multirow{6}{*}{Test}
    &6.0$\times$8.0$\times$3.8
    &\multirow{2}{*}{(2.0, 4.0, 1.65)} &\multirow{2}{*}{0.6, 1.5, 2.4, 3.3}
    &0.2: 0.2: 0.8 &5 &\multirow{2}{*}{21, 3, 40}\\
    &(Large) & & &0.6 &-5: 5: 15
    &\vspace{0.08cm} \\
    &5.0$\times$7.0$\times$3.0
    &\multirow{2}{*}{(2.5, 3.0, 1.5)}  &\multirow{2}{*}{0.7, 1.4, 2.1}
    &0.2: 0.2: 0.8 &5 &\multirow{2}{*}{8, 9, 33}\\
    &(Medium) & & &0.6 &-5: 5: 15
    &\vspace{0.08cm}\\
    &4.0$\times$4.0$\times$2.7
    &\multirow{2}{*}{(1.8, 1.7, 1.6)}  &\multirow{2}{*}{0.8, 1.3}
    &0.2: 0.2: 0.8 &5 &\multirow{2}{*}{19, 20, 27}\\
    &(Small) & & &0.6 &-5: 5: 15 \\
    \hline
    \hline
  \end{tabular}
\end{table*}

In this section, the performance of the proposed method is measured on both simulated and real-world data. We first describe the details of the experimental setup, and then give the experimental results and discussions.

\subsection{Experimental Setup}

\begin{figure}[t]
  \centering
  \includegraphics[width=0.85\linewidth]{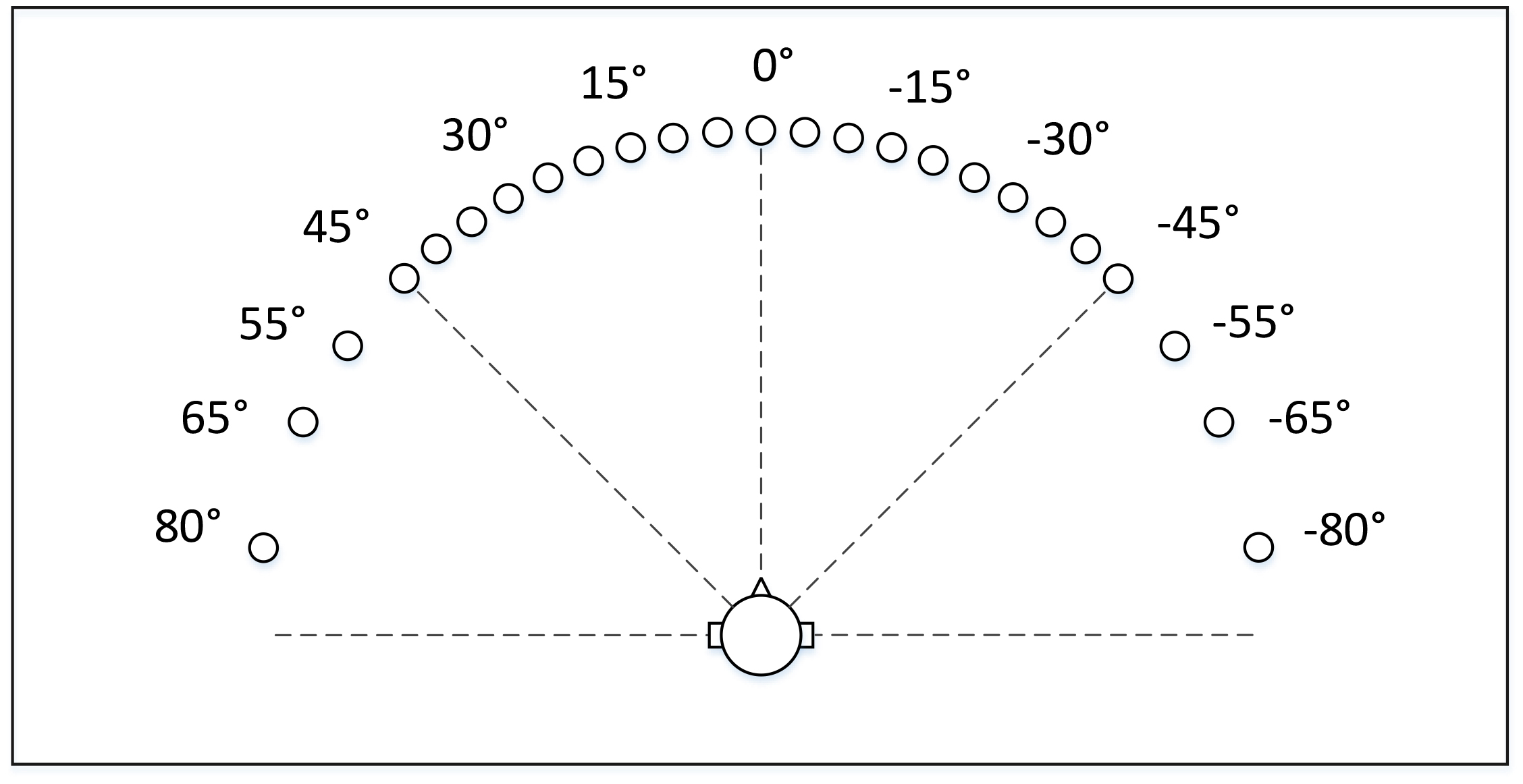}
  \caption{Illustration for the candidate directions of the sound source relative to binaural microphones.}
  \label{fig:exp_head}
\end{figure}

\subsubsection{Simulated data}

Fifteen room configurations are simulated using the image method \cite{RIR79} implemented by the Roomsim toolbox \cite{Roomsim05}. The data generation configurations are summarized in Table~\ref{tab:room}, among which ten room settings are used for training, two for validation and three for test. All the experiments are carried out using binaural microphones with prominent shadow effect. The speech sound source is located in the same horizontal plane as the two microphones, and the candidate source directions are [-80$^\circ$, -65$^\circ$, -55$^\circ$, -45$^\circ$: 5$^\circ$: 45$^\circ$, 55$^\circ$, 65$^\circ$, 80$^\circ$], as illustrated in Fig.~\ref{fig:exp_head}. The acoustic impulse response or binaural room impulse response (BRIR) is generated using the Roomsim toolbox and the HRIRs from the CIPIC database \cite{HRIR_CIPIC01}. The CIPIC database contains HRIRs of 45 different subjects including 27 male human subjects, 16 female human subjects, as well as two KEMAR dummy heads \cite{HRIR_KEMAR95} with either large or small pinnaes. These head subjects are distinctively used for training, validation and test, as shown in Table~\ref{tab:room}. We randomly select speech recordings from TIMIT dataset \cite{TIMIT93}, and truncate each to obtain speech segments with a duration of 0.5 s. These segments are divided into three parts to act as source signals for training, validation and test, respectively. The sensor signals are created by convolving the BRIRs with the source signals. We use the white, babble and factory noise signals from the NOISEX-92 database \cite{Noise92}. Each noise signal is split as training, validation and test segments, respectively, without overlap between them. With these noise signals, the arbitrary noise field generator \cite{ANF} is employed to generate a binaural diffuse noise field \cite{Diffuse08} according to the microphone distance of corresponding binaural setups. Diffuse noise is scaled and added to each sensor signal with different signal-to-noise ratios (SNRs), in order to simulate the acoustic conditions with various levels of noise. When generating each instance, the source signal, noise signal, RT$_{60}$ and SNR are randomly given within the aforementioned settings.
The numbers of instances for training, validation and test are 120,000 and 24,000, 324,000 respectively.

\begin{figure}[t]
  \centering
  \includegraphics[width=0.85\linewidth]{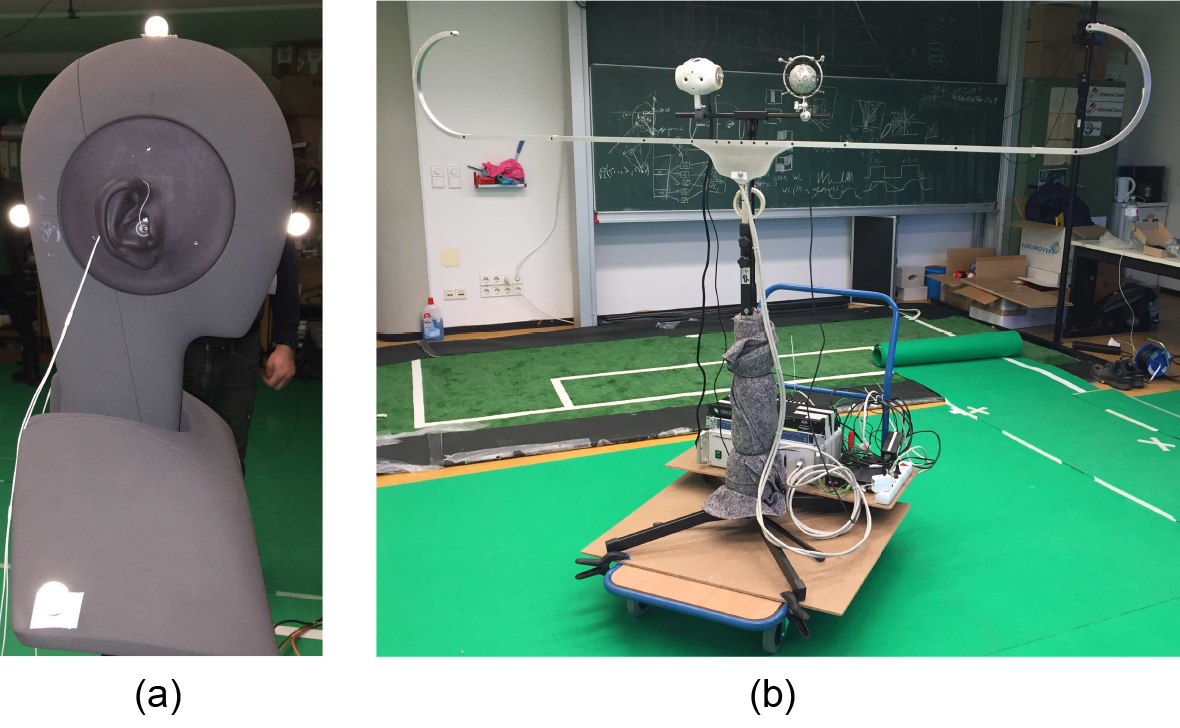}
  \caption{The settings for LOCATA data corpus \cite{LOCATA18}. (a) The hearing aids and dummy head. (b) The recording environment. }
  \label{fig:exp_locata_scene}
\end{figure}

\subsubsection{Real-world data}
The localization and tracking (LOCATA) challenge data corpus \cite{LOCATA18} provides audio signals recorded in the computing laboratory of the Department of Computer Science at the Humboldt University Berlin as shown in Fig.~\ref{fig:exp_locata_scene}. The room size is 7.1 m $\times$ 9.8 m $\times$ 3 m with a reverberation time of 0.55 s. The source utterances from the Centre for Speech Technology Research Voice Cloning ToolKit dataset \cite{VCTK17} are played back by static loudspeakers or read live by five moving human talkers. Two hearing aids are separately equipped at the left and the right ears of the dummy head with a distance of 157 mm.  Each hearing aid has two microphones with an inter-microphone distance of 9 mm. The signals captured by microphone 1 and 3 are used for evaluation in this work. We consider task 3 in which the speaker is moving  in the frontal azimuthal half-plane of the static head.

\subsubsection{Parameter setting}
The sampling rate of binaural signals is 16 kHz. STFT is performed with a window length of 32 ms and a frame shift of 16 ms.  Since the energy of speech signals mainly lies in the range from 0 to 4 kHz, signals in this frequency range are used for localization, and correspondingly the number of frequencies $F$ is 128. The maximum IID value $\Delta I_{\rm{max}}$ is set to 20.  Each test signal segment is with a duration of 0.5 s, and correspondingly the number of time frames $N$ is 31, unless otherwise stated. During training, we train the monaural enhancement network first, and then train the DP-RTF learning network with the monaural enhancement network frozen. The model is trained using the Adam optimizer.

\subsubsection{Evaluation metrics}

For the simulated data, the performance of DOA estimation is evaluated using two metrics:
(i) Localization accuracy (ACC), calculates the percentage of the correctly localized instances in all test instances. The prediction is considered to be correct if the dictionary matching result of Eq.~\eqref{eq_SSL} is the correct one.
(ii) Mean absolute error (MAE), is the averaged absolute error between the estimated and the ground-truth DOAs over all test instances.
For the real-world data, we adopt the metrics provided by the LOCATA dataset to evaluate the performance of DOA estimation so as to facilitate the comparison with other methods present in the challenge. The metrics include the MAE of azimuth estimation for the successfully localized source, the probability of source detection (PD) and the false alarm rate (FAR), only for the voice-active periods \cite{LOCATA_Results20}.

\subsubsection{Comparison method}
To verify the effectiveness of the proposed method, four deep learning based localization methods are compared.

DOA-CNN \cite{CNN19}:
The architecture of this method is with one convolutional layer and three FC layers. The input vector is the phase of the STFT coefficient of single-frame sensor signals. This model outputs the posterior probability for one single time frame, and the DOA is determined by taking the average of the posterior probabilities of multiple frames.

Mask-GCC \cite{Mask_RNN19}: It uses a two-layer bi-directional long short-term memory (BLSTM) model to predict the phase-sensitive mask (PSM) from the single-channel log magnitude spectrogram. The predicted mask is used to weight the TF components when computing the GCC-phase transform (PHAT) or the steered-response SNR. We will only present the results of GCC-PHAT in this experiment, as it was shown in our preliminary experiments that GCC-PHAT averagely outperforms steered-response SNR.

IPD-EN \cite{IPD19}: The sine and cosine of single-frame full-band IPD are taken as the localization feature. It uses four FC layers to predict the clean localization feature from the contaminated ones. The DOA estimation of this work was originally designed for regular microphone array. To make it feasible for binaural data, we modified it to the feature matching technique similar to Eq.~\eqref{eq_SSL}.

RTF-EN \cite{YBCAAI21}: It is our previous work. The network contains four FC layers. It takes multi-frame contaminated DP-RTF as input feature, and predicts single-frame clean one.

\subsection{Ablation Experiments}

\subsubsection{DP-RTF regression versus location classification}

\begin{table}[t]
  \caption{Performance (ACC, MAE)  for DOA estimation with different DOA estimation methods} 
  \label{tab:loss}
  \centering
  \renewcommand\arraystretch{1.05}
  \tabcolsep0.04in
  \begin{tabular}{ccccccccccccc}
    \hline
    \hline
    DOA estimation method &Array &Test head &ACC [\%] & MAE [$^{\circ}$] \\
    \hline
    \multirow{4}{*}{Location classification}
    &\multirow{2}{*}{Matched} &21 &\textbf{89.7}  &2.1\\
    & &40 &\textbf{86.1} &2.3\\
    &\multirow{2}{*}{Mismatched} &21 &49.4 &5.3\\
    & &40 &35.5 &\ 7.4 \vspace{0.08cm} \\
    &\multirow{2}{*}{Matched}  &21 &\textbf{89.7} &\textbf{1.8} \\
    \textbf{DP-RTF regression} & &40 &83.1 &\textbf{2.2} \\
    \textbf{(prop.)} &\multirow{2}{*}{Mismatched}  &21 &\textbf{89.3} &\textbf{2.0} \\
    &  &40 &\textbf{80.0} &\textbf{2.7} \\
    \hline
    \hline
    \end{tabular}
\end{table}

We compare the performance of DOA estimation when treating it as either a DP-RTF regression  problem (the proposed one) or a location classification problem. When classifying locations, one FC layer is added at the end of the DP-RTF learning network to output the posterior probability of each source direction, and the training loss is cross entropy. This location classification setup is similar to many deep-classification-based sound source localization methods \cite{DNN15, DNN16, CCF17, CNN19, Mask_DOA19, CRNN_intensity19,SCL_CNN20, ETE_Ma19, CRNN18}, which divides the localization space into a number of portions and each portion corresponds to one location class. The default array-mismatched setup mentioned in Section V-A 1) uses different binaural arrays for training and test, for which the DP-RTF learning (and thus DOA estimation) should generalize across different arrays. To analyze the characteristics of array generalization, we also test the array-matched condition, namely the same head is used for training and test. For the array-matched case, head 21 and 40 are used for test, and the training data is generated using the same heads with the acoustic settings listed in Table~\ref{tab:room}.

Table~\ref{tab:loss} shows the results for different DOA estimation methods under various array conditions.
It can be seen that compared with location classification, DP-RTF regression achieves worse localization accuracy results but better MAE results for the array-matched case.
The location classification setup makes the feature space of DOA classes as separate as possible to pursue a high classification accuracy, but a wrong classification may correspond to a large DOA estimation error. By contrast, the DP-RTF regression setup estimates the DP-RTF in a continuous feature space, with the aim to minimize the DP-RTF regression error and thus the DOA estimation error. For the array-mismatched case, DP-RTF regression outperforms location classification by a large margin. The feature-to-location relations are distinct for different arrays, which is problematic for the location classification setup to learn an one-to-one feature (or signal)-to-location mapping. This problem is handled in this work by explicitly adopting the array-specific feature dictionaries for DOA estimation based on look-up as in Eq.~\eqref{eq_SSL}.

For the proposed DP-RTF regression setup, a performance degradation from the array-matched case to the array-mismatched case can be observed.  The performance degradation is insignificant for head 21, while notable for head 40. This indicates that the DP-RTF space of one head (like head 21) indeed can be well approximated by one of the other training heads. However, the diversity of training heads should be further increased to cover more unseen heads (like head 40).

\begin{table}[t]
 \caption{Performance (ACC, MAE) For DOA estimation with and without magnitude-related model}
  \label{tab:mag}
  \centering
  \renewcommand\arraystretch{1.05}
  \tabcolsep0.12in
  \begin{tabular}{cccccccccccccc}
    \hline
    \hline
    Nework  & ACC [\%]  & MAE [$^{\circ}$]\\
    \hline
    \multirow{1}{*}{W/o magnitude-related model}  &75.4 &5.0 \\
    \multirow{1}{*}{\textbf{W/ magnitude-related model (prop.)}} &\textbf{85.4} &\textbf{2.2}  \\
    \hline
    \hline
   \end{tabular}
\end{table}

\subsubsection{Contribution of magnitude-related model}
To evaluate the contribution of the magnitude-related parts of the proposed method, we also test the performance of the proposed model with the magnitude-related parts being removed, in which the network input and output are the dual-channel phase spectrograms and the phase part of DP-RTF, respectively. The experimental results are shown in Table~\ref{tab:mag}. It can be observed that the magnitude-related model brings a 10\% increase on ACC and a 2.8$^\circ$ decrease on MAE, which confirms that the spatial and spectral cues encoded in magnitude are crucial for binaural sound source localization.

\begin{table}[t]
 \caption{Performance (ACC, MAE) for DOA estimation using different DP-RTF learning network architectures}
  \label{tab:DPL}
  \centering
  \renewcommand\arraystretch{1.05}
  \tabcolsep0.03in
  \begin{tabular}{cccccccccccccc}
    \hline
    \hline
    Network architecture &CNN depth & ACC [\%]  & MAE [$^{\circ}$] \\
    \hline
    \multirow{1}{*}{Joint CRNN} & \multirow{1}{*}{10}
    &63.7 &6.1 \\
    \multirow{1}{*}{(Not-branched) CNN + joint CRNN} & \multirow{1}{*}{12}
    &65.6 &5.8 \\
    \multirow{1}{*}{\textbf{Branched CNN + joint CRNN (prop.)}} &\multirow{1}{*}{12}
    &\textbf{67.5} &\textbf{5.4} \\
    \hline
    \hline
   \end{tabular}
\end{table}
\subsubsection{Network architecture comparison for DP-RTF learning}
Some preliminary experiments have been done to determine the architecture and parameters of the CRNN for DP-RTF learning. Eventually, the proposed architecture consists of some branched CNN layers, followed by a joint CRNN module. To show the effectiveness of CNN branches, we also report the performance of the joint CRNN module solely. For fair comparison, the proposed architecture is also compared with the architecture which adds one extra network with several (not-branched) CNN layers at the beginning of the joint CRNN module. The two architectures own an identical depth of CNN layers.

Table~\ref{tab:DPL} shows the results of the three network architectures. The results are averaged over the following conditions: the source-to-array distances are 3.3 m, 2.1 m and 1.3 m for the large, medium and small rooms respectively, and the (RT$_{60}$, SNR) conditions are (0.6 s, 5 dB), (0.6 s, 0 dB), (0.6 s, -5 dB) and (0.8 s, 5 dB). It can be seen that the proposed network architecture outperforms the others by a noticeable margin. The improvement by CNN branches is mainly attributed to the fact that it separately extracts the magnitude and phase patterns and meanwhile highlights the patterns of reliable TF regions.

\subsubsection{Influence of monaural enhancement network}

\begin{table}[t]
 \caption{Performance (SDR, IID error, IPD error) of monaural speech enhancement}
  \label{tab:enhance}
  \centering
  \renewcommand\arraystretch{1.05}
  \tabcolsep0.18in
  \begin{tabular}{cccccccccccccc}
    \hline
    \hline
    Sensor signal  &SDR [dB] &IID error &IPD error\\
    \hline
    \multirow{1}{*}{Unprocessed}  &-7.4 &\textbf{0.26} &1.62\\
    \multirow{1}{*}{Enhanced} &\textbf{2.1} &0.33 &\textbf{1.59} \\
    \hline
    \hline
   \end{tabular}
\end{table}

\begin{table}[t]
 \caption{Performance (ACC, MAE) for DOA estimation with or without monaural speech enhancement}
  \label{tab:addenhance}
  \centering
  \renewcommand\arraystretch{1.05}
  \tabcolsep0.1in
  \begin{tabular}{cccccccccccccc}
    \hline
    \hline
    Network  & ACC [\%] & MAE [$^{\circ}]$\\
    \hline
    \multirow{1}{*}{Unprocess.}
    &16.1 &29.4\\
    \multirow{1}{*}{Monaural-enhance.}
    &20.2 &26.4  \\
    \multirow{1}{*}{DP-RTF-learn.}
    &67.5 &5.4  \\
    \multirow{1}{*}{\textbf{Monaural enhance. + DP-RTF learn.}}
    &\textbf{69.1} &\textbf{5.0}\\
    \hline
    \hline
   \end{tabular}
\end{table}

To evaluate the influence of monaural speech enhancement on sound source localization, we first test the performance of monaural speech enhancement for noise and reverberation removal and binaural cues recovery.
Table~\ref{tab:enhance} shows the performance of monaural speech enhancement in terms of signal-to-distortion ratio (SDR), IID error and IPD error. The DP-RTF are directly estimated from the magnitude and phase spectrograms of the unprocessed or enhanced sensor signals by taking the ratio of STFT coefficients between two microphones. The IID error and the IPD error refer to the MSE of the estimated DP-RTFs (in the real-valued form of Eq.~\eqref{eq_DP_RTF_vec}) relative to the ground truth, which are averaged over the voice-active TF bins of all test instances. The SDR improvement from the unprocessed signals to the enhanced signals is about 9.5 dB, which means the monaural enhancement method can largely remove the noise and reverberation. However, the IID error and IPD error are not well reduced as originally expected. This is possibly because that the noise and reverberation reduction is mainly performed on the speech-inactive TF bins.

We then test the performance of DOA estimation without and with the monaural speech enhancement. The experimental results are shown in Table~\ref{tab:addenhance}. The evaluation data is the same as that in the Section V-B 3). The unprocess. and monaural-enhance. methods use the unprocessed and the speech-enhanced spectrograms to compute the DP-RTF, respectively. The DP-RTF-learn. method takes the unprocessed spectrograms of the binaural signals as input. As for the monaural enhance. + DP-RTF learn. method, the enhanced and the unprocessed spectrograms are stacked along the microphone channel dimension.
It can be seen that the monaural-enhance. method improves the performance to some degree when compared with the DOA estimation with the unprocessed spectrograms. But it still performs poorly, since the localization features contaminated by noise and reverberation are not well recovered by monaural speech enhancement.
Stacking the enhanced and unprocessed spectrograms as the input of DP-RTF learning network is able to improve the performance (relative to inputting solely the unprocessed spectrograms). This indicates that, although better localization features cannot be extracted directly from the enhanced spectrograms, the enhanced spectrograms can still provide some useful information, such as the TF-wise speech activities.

\subsection{Comparison with Other Methods}

\begin{figure}[t]
  \centering
  \includegraphics[width=1\linewidth]{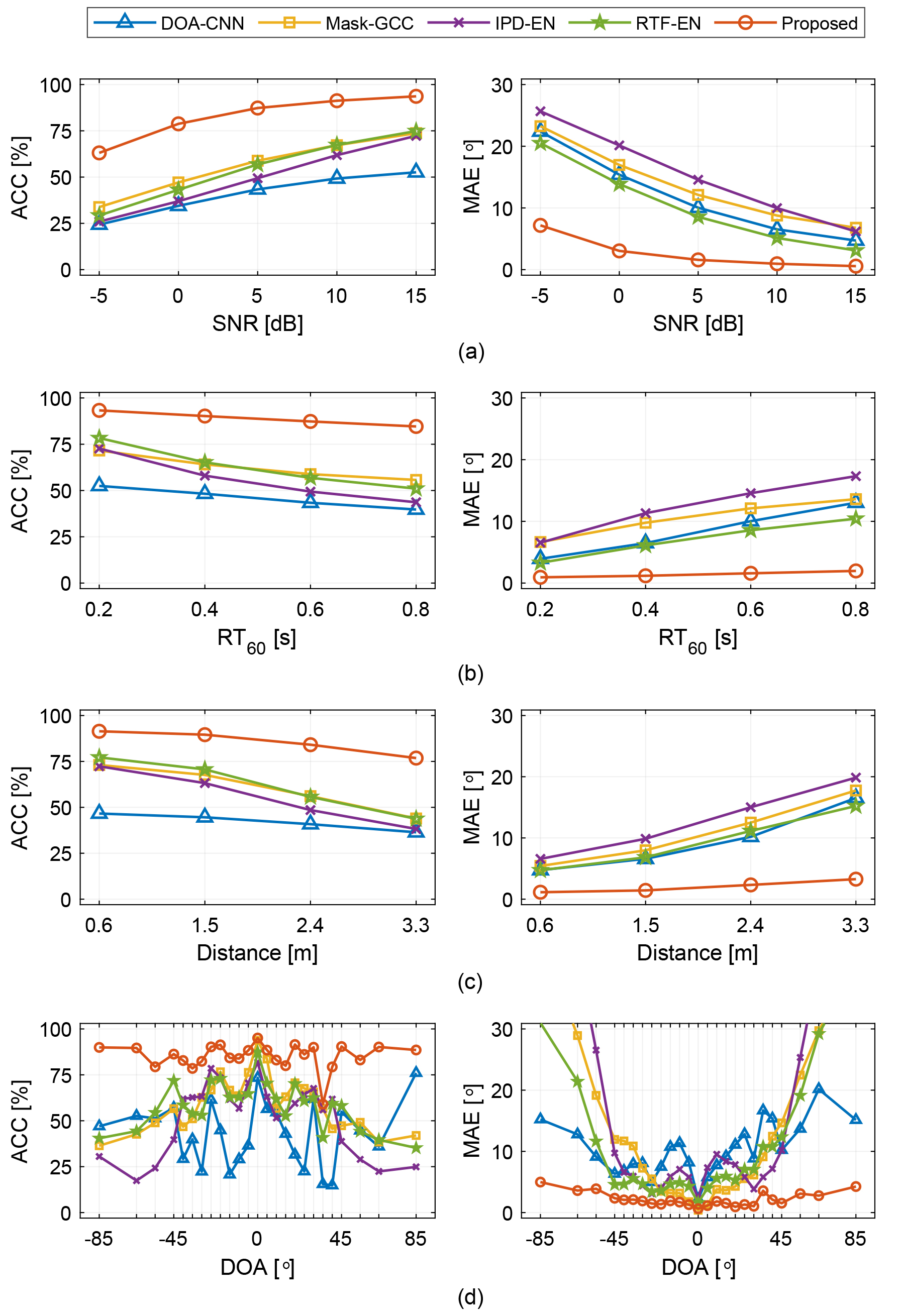}
  \caption{Performance (ACC, MAE) of different localization methods in the simulated scenarios with (a) different SNRs (RT$_{60}$ = 0.6 s), (b) different RT$_{60}$s (SNR = 5 dB), (c) different source-to-array distances (large room), and (d) different DOAs.}
  \label{fig:exp_sim_com}
\end{figure}

\subsubsection{Evaluation with simulated data}

The proposed method is compared with four deep learning based methods, namely DOA-CNN \cite{CNN19}, Mask-GCC \cite{Mask_RNN19}, IPD-EN \cite{IPD19} and RTF-EN \cite{YBCAAI21}. All the five methods are trained with the same simulated data. The ACCs and MAEs of these five methods under different acoustic conditions are shown in Fig.~\ref{fig:exp_sim_com}. It can be observed that the proposed method largely outperforms all the comparison methods under all the test conditions. The DOA-CNN method neglects the difference of the feature-to-location mapping for different binaural arrays, and thus shows a poor generalization ability to unseen binaural arrays. Both the IPD-EN and RTF-EN methods recover clean spatial features from the contaminated ones, and the information of speech spectra is not used. The Mask-GCC method takes full advantage of spectral cues to predict the TF mask so that more weights are placed on the TF bins dominated by the target speech. However, the Mask-GCC method is highly dependent on the accuracy of the mask estimation especially under adverse acoustic conditions. By contrast, the proposed method utilizes a well designed CRNN network to estimate the clean localization feature by automatically extracting and revising the spatial feature from the binaural spectrograms, and meanwhile by leveraging the spectral information to aid the spatial feature estimation.

When the acoustic conditions become worse, namely the level of noise and reverberation increases, the performance of all these methods degrade. Their performance also degrades with the increasing of the source-to-array distance, due to the decreased energy ratio of direct-path sound to reverberation. The proposed method shows a relatively smaller performance degradation with the changing of these factors, which indicates a stronger robustness against noise and reverberation. All the methods achieve a comparable performance when the source is at 0$^{\circ}$. The performance of Mask-GCC, IPD-EN and RTF-EN shows an apparent decreasing when the DOA is approaching to the end-fire directions, while the proposed method seems to be less sensible to the source direction. The difficulty of localizing end-fire directions is possibly due to the highly wrapped IPDs caused by large time delay of arrivals. The good performance of the proposed method for the end-fire directions indicates that the proposed method is able to recover the full-band IPD wrapping structure, which can be testified by the DP-RTF estimation examples shown in Fig.~\ref{fig:exp_rtf_doa_meth}.

\subsubsection{Evaluation with real-world data}

\begin{table}[t]
  \caption{Performance (MAE, PD, FAR) comparison of different localization methods on task 3 of the LOCATA challenge}
  \label{tab:sota_real}
  \centering
  \renewcommand\arraystretch{1.05}
  \tabcolsep0.02 in
  \begin{tabular}{cccccccccccc}
    \hline
    \hline
    \multirow{2}{*}{Method}  &\multicolumn{3}{c}{Error tolerance: 30$^{\circ}$} & &\multicolumn{3}{c}{Error tolerance: 10$^{\circ}$}\\
    \cline{2-4}  \cline{6-8}
    &MAE [$^{\circ}$]  &PD [\%] &FAR [/s] & &MAE [$^{\circ}$]  &PD [\%] &FAR [/s]\\
    \hline
    MUSIC \cite{LOCATA_MUSIC}   &16.0 &- &-     & &- &- &-\\
    CIMPL \cite{LOCATA_CIMPL}   &7.2 &- &-      & &- &- &-\\
    CTF-DPRTF \cite{LXF19}      &4.2 &98.1 &2.3 & &3.5 &90.4 &11.6\\
    DOA-CNN \cite{CNN19}        &3.3 &97.8 &2.7 & &2.5 &92.7 &8.8\\
    Mask-GCC \cite{Mask_RNN19}  &4.5 &98.9 &1.3 & &2.9 &86.6 &16.2\\
    IPD-EN \cite{IPD19}         &3.1 &94.9 &6.2 & &2.6 &91.6 &10.2\\
    RTF-EN \cite{YBCAAI21}      &3.3 &99.4 &0.8 & &2.9 &95.5 &5.5\\
    \textbf{Proposed}           &\textbf{2.2} &\textbf{99.5} &\textbf{0.6} & &\textbf{2.1} &\textbf{98.8} &\textbf{1.5}\\
    \hline
    \hline
    \end{tabular}
\end{table}

The proposed method is compared with the above four comparison methods on the real-world data. The deep learning-based models trained with simulated data are directly  tested on the real-world data. The evaluation set of task 3 from the LOCATA dataset is utilized. Only the signals recorded by microphone 1 and 3 are used to perform binaural localization. Since the HRIRs for the LOCATA dataset are unavailable, we use the averaged DP-RTF of all the CIPIC HRIRs to form the DP-RTF dictionary of the test array for Mask-GCC, IPD-EN, RTF-EN and the proposed method. In addition, the methods presented in the LOCATA challenge are also compared, including MUSIC \cite{LOCATA_MUSIC}, CIMPL \cite{LOCATA_CIMPL} and CTF-DPRTF \cite{LXF19}.
The sound source is considered to be localized successfully if the azimuth error is not larger than a error tolerance. The error tolerance is set to 30$^{\circ}$ and 10$^{\circ}$, respectively. For the MUSIC and CIMPL methods, the results presented in \cite{LOCATA_Results20} with a error tolerance of 30$^{\circ}$ are directly quoted here. Note that these results are obtained by using all the four microphones.

\begin{figure}[t]
  \centering
  \includegraphics[width=1\linewidth]{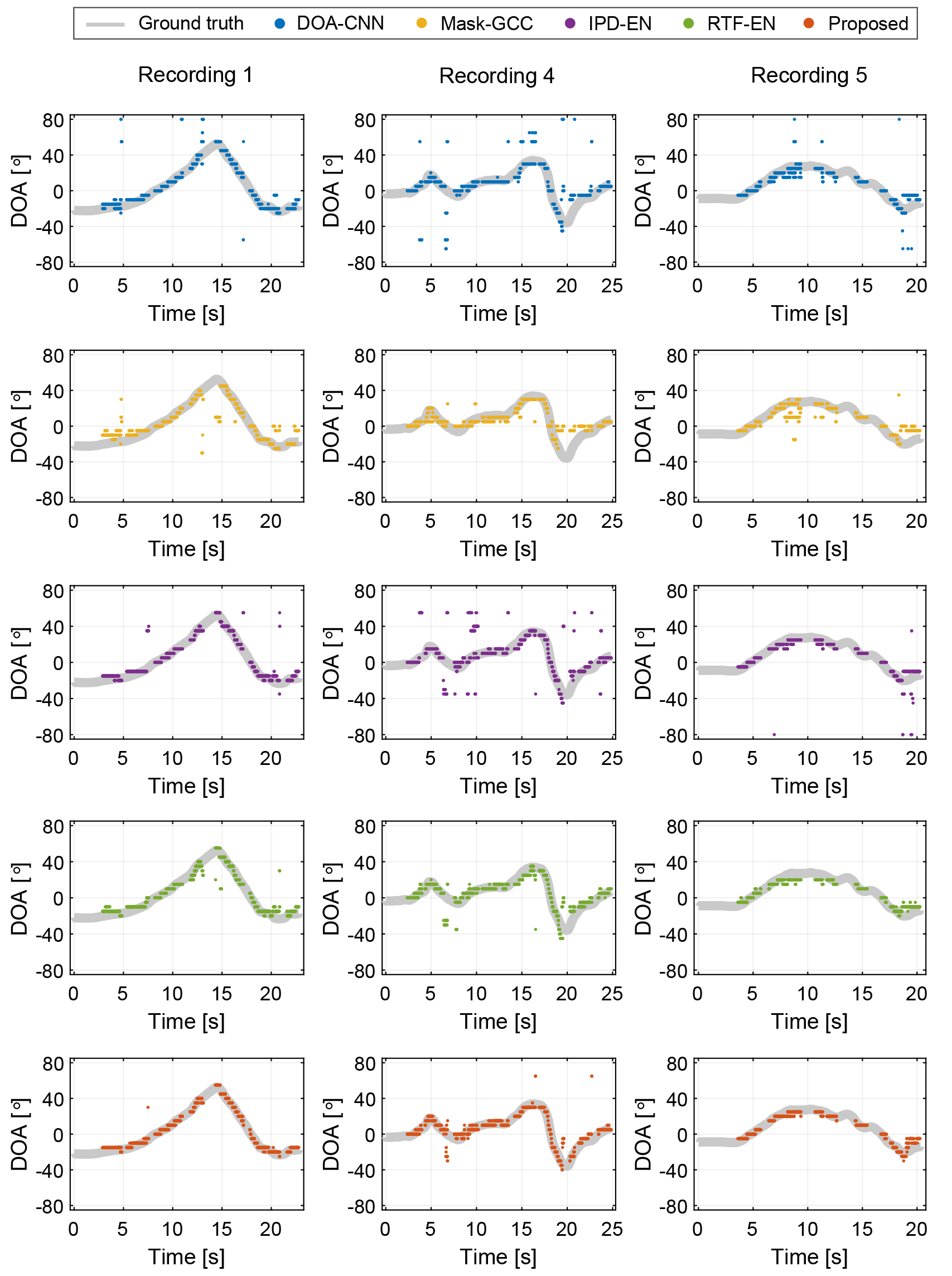}
  \caption{Illustration of DOA estimation for a moving source from task 3 of the LOCATA challenge, using DOA-CNN, Mask-GCC, IPD-EN, RTF-EN and the proposed method respectively.
  Only the estimates during voice-active periods are shown.}
  \label{fig:exp_locata}
\end{figure}

Table~\ref{tab:sota_real} shows the performance comparison in terms of MAE, PD and FAR. It can be observed that the proposed method outperforms other methods on all the three evaluation metrics. When the error tolerance drops from 30$^{\circ}$ to 10$^{\circ}$, the superiority of our method seems more prominent in terms of PD and FAR. It manifests that the proposed DP-RTF learning network can well generalize to the unseen binaural array and the real-world acoustic conditions. Some localization examples of the five deep learning based methods are illustrated in Fig.~\ref{fig:exp_locata}. It is apparent that DOA-CNN and IPD-EN provide a larger number of erroneous DOA estimates with obvious deviation from the ground truths, which indicates a larger probability of miss detection or false alarm. The proposed method achieves the lowest error of DOA estimation, especially when the source takes a turn around.

\subsection{DP-RTF Visualization}

\begin{figure}[t]
  \centering
  \includegraphics[width=1\linewidth]{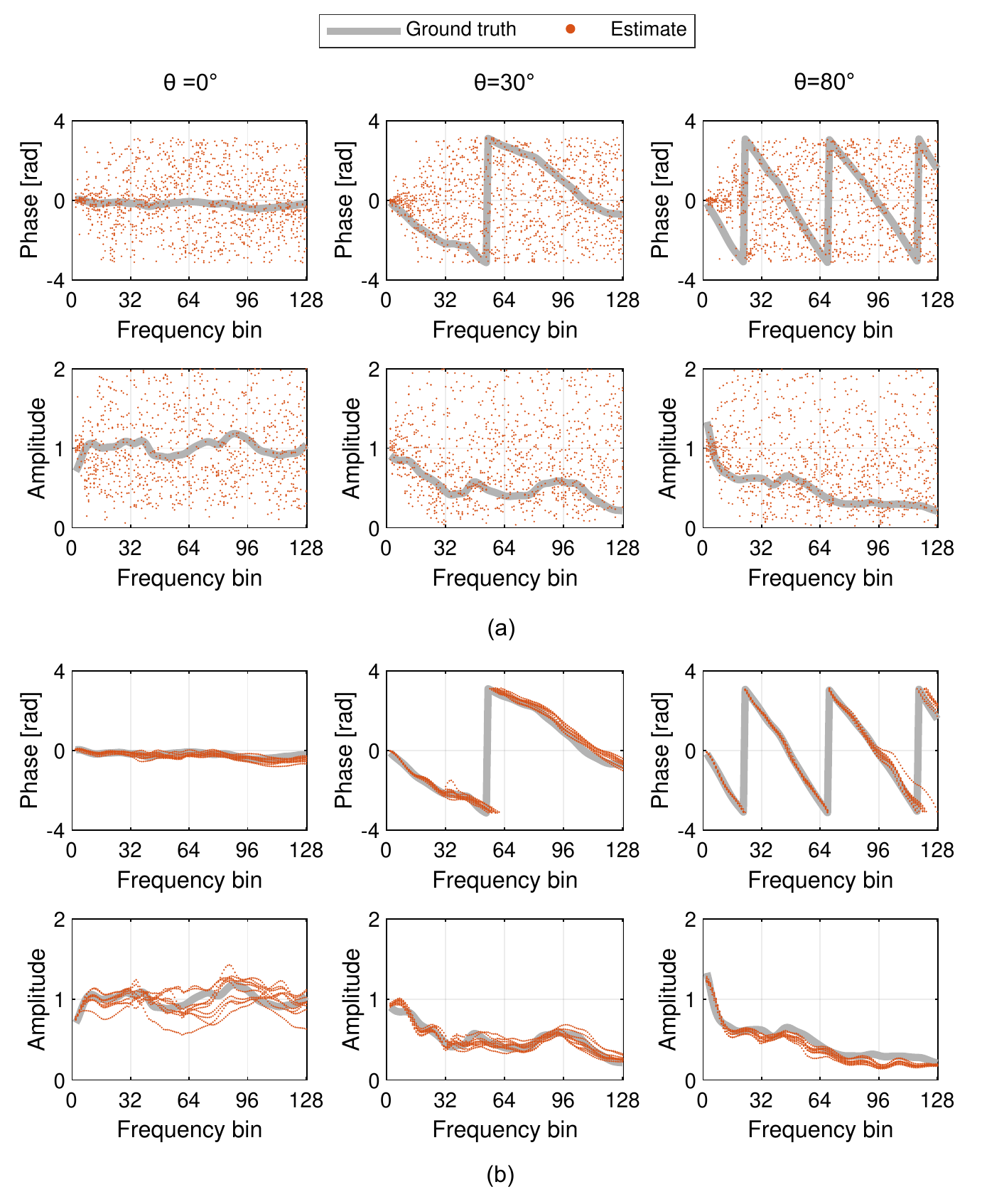}
  \caption{Phase and amplitude of the estimated DP-RTFs for different source directions.
  (a) is estimated directly from the STFT coefficients of original binaural signals.
  (b) is estimated by the proposed method. Ten trials are conducted and presented for each direction.
  The acoustic conditions of this experiment are: Medium room, head 21, RT$_{60}$ = 0.6 s, SNR = 5 dB (babble noise) and source-to-array distance = 2.1 m. }
  \label{fig:exp_rtf_doa_meth}
\end{figure}

\begin{figure}[t]
  \centering
  \includegraphics[width=1\linewidth]{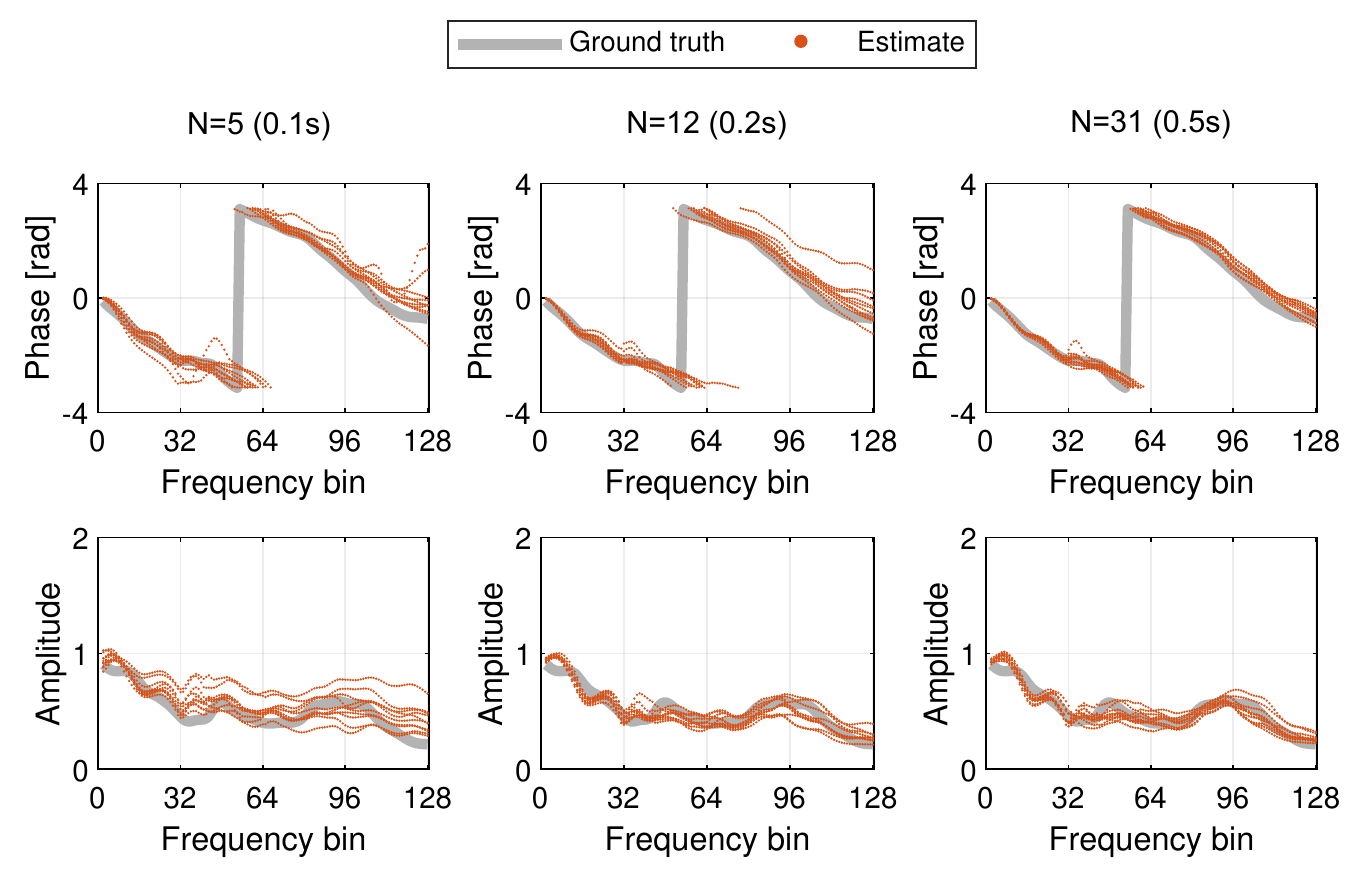}
  \caption{Phase and amplitude of the estimated DP-RTFs using sensor signals with different time durations. Ten trials are conducted and presented for each direction. The acoustic conditions of this experiment are: Medium room, head 21, $\theta$ = 30$^{\circ}$, RT$_{60}$ = 0.6 s, SNR = 5 dB (babble noise) and source-to-array distance = 2.1 m.}
  \label{fig:exp_rtf_nt}
\end{figure}

The network output of the proposed method can be interpreted physically, and thereby facilitates us to do factor analysis and method comparison by visualizing the DP-RTF features. In Fig.~\ref{fig:exp_rtf_doa_meth}, the phase and amplitude of DP-RTF are plotted along with frequency bins for different source directions. With the increasing of frequency, phase tends to be wrapped for the source at 30$^\circ$ and 80$^\circ$. The DP-RTF phase and amplitude estimated from original contaminated signals are scattered, while that estimated by the proposed method are clustered around the ground-truth line. It confirms the proposed method can well recover the direct-path spatial cues and meanwhile reduce the effect of noise and reverberation. Fig.~\ref{fig:exp_rtf_nt} shows the DP-RTF predictions when taking as input different lengths of binaural signals. It can be observed that with the increasing of time duration, the phase and amplitude estimation become more concentrative around the ground-truth line. This demonstrates that a larger signal duration, which provides more temporal context, is required for reliable DP-RTF estimation. However, a larger signal duration will lead to a longer estimation latency, which is problematic for real-time moving speaker localization. Therefore, the signal duration for one DP-RTF estimation is set to 0.5 s as a good trade off in this work, and thus a good performance can be achieved for the moving speaker scenarios, as shown in Fig.~\ref{fig:exp_locata}.

\section{Conclusions}
This paper makes full use of the spatial and spectral information to learn DP-RTF for binaural sound source localization under adverse acoustic conditions. The DP-RTF learning network follows a separate-to-joint learning process to embed the sensor signals into a real-valued direct-path feature, in which way the spatial cues are preserved and the distortion caused by noise and reverberation is suppressed.
The monaural enhanced speech is added before the DP-RTF learning network to aid the learning of inter-channel feature. The network trained with many simulated binaural arrays shows favorable generalization ability when test on unseen binaural arrays. Experiments conducted on both simulated and real-world data verify the advantage of our method over several other methods for sound source localization in scenarios with different levels of noise and reverberation, and source-to-array distances.

The monaural speech enhancement is able to largely reduce noise and reverberation, but possibly makes the spatial cues distorted. In the future work, a binaural speech enhancement method with spatial cues preserved can be investigated to provide more helpful information for localization.
In this work, we study the feature regression based DOA estimation method for a single source. In future work, this will be extended to the case of multiple sources by simultaneously predicting the DP-RTFs of multiple sources. Currently, the proposed method can localize one moving source by block-wisely estimating the time-varying DP-RTF feature. We may revise the proposed network to better model the dynamic temporal context for sound source tracking in the future.


%





\ifCLASSOPTIONcaptionsoff
  \newpage
\fi



\bibliographystyle{IEEEtran}
\bibliography{mybib}
%
%
%

%
\begin{IEEEbiography}[{\includegraphics[width=1in,height=1.25in,clip,keepaspectratio]{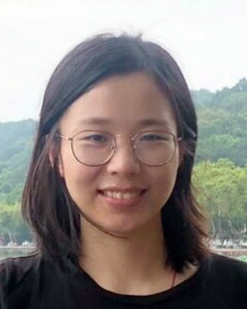}}]{Bing Yang}
received the B.E. degree in automation from University Of Science and Technology Beijing, Beijing, China, in 2015. She is currently working toward the Ph.D. degree in the School of Electronics Engineering and Computer Science, Peking University, Beijing, China. Her current research interests include  multimicrophone speech and audio processing for sound source localization and tracking.
\end{IEEEbiography}

\vfill

\begin{IEEEbiography}[{\includegraphics[width=1in,height=1.25in,clip,keepaspectratio]{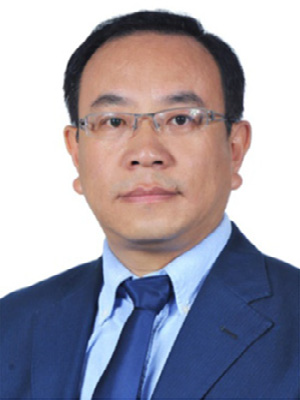}}]{Hong Liu}
received the Ph.D. degree in mechanical electronics and automation in 1996. He serves as a Full Professor in the School of EE$\&$CS, Peking University (PKU), China. Prof. Liu has been selected as Chinese Innovation Leading Talent supported by National High-level Talents Special Support Plan since 2013. He is also the Director of Open Lab on Human Robot Interaction, PKU, his research fields include computer vision and robotics, image processing, and pattern recognition. Dr. Liu has published more than 200 papers
and gained the Chinese National Aerospace Award, Wu Wenjun Award on Artificial Intelligence, Excellence Teaching Award, and Candidates of Top Ten Outstanding Professors in PKU. He is an IEEE member, vice president of the Chinese Association for Artificial Intelligent (CAAI), and vice-chair of the Intelligent Robotics Society of CAAI. He has served as keynote speakers, co-chairs, session chairs, or PC members of many important international conferences, such as IEEE/RSJ IROS, IEEE ROBIO, IEEE SMC, and IIHMSP. Recently, Dr. Liu publishes many papers on international journals and conferences such as Pattern Recognition, IEEE Transactions on Image Processing, and International Joint Conference on Artificial Intelligence, the field involves path planning, action recognition, and person re-identification.
\end{IEEEbiography}

\begin{IEEEbiography}[{\includegraphics[width=1in,height=1.25in,clip,keepaspectratio]{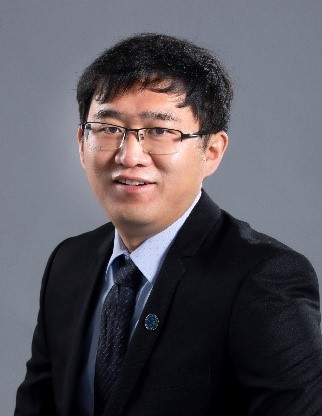}}]{Xiaofei Li} is an assistant Professor at Westlake University, China. Before, he worked at INRIA Grenoble Rh\^{o}ne-Alpes, France, as a post-doctoral researcher from Feb. 2014 to Jan. 2016, and as a starting research scientist from Feb. 2016 to Dec. 2019. He obtained his PhD degree from Peking University, China, in Jul. 2013. His research interests lie in the field of acoustic, audio and speech signal processing, including the topics of speech denoising, dereverberation, separation and localization; sound/speech semi-supervised learning and unsupervised pre-training; sound field reproduction and personal sound zone.
\end{IEEEbiography}

\vfill



\end{document}